\documentclass[12pt]{article}
\usepackage{amsmath}
\usepackage{wick}
\usepackage{amssymb}
\usepackage[dvips]{graphicx}
\usepackage{amsmath,amscd}
\usepackage{color}
\usepackage{mathrsfs}
\usepackage{calligra}
\usepackage{latexsym}
\usepackage{cite}
\makeatletter

  \@addtoreset{equation}{section}
\makeatother
\usepackage[dvips]{graphicx}

\newcommand{\LBox}{\mbox{\Large$\Box$}}

\begin{document}
\begin{titlepage}
\thispagestyle{empty}
\begin{flushright}

\end{flushright}
\vspace{3mm}

\begin{center}
{\huge Radion stabilization in the presence of\\ \vspace{3mm}  Wilson line phase}
\end{center}

\begin{center}
\lineskip .45em
\vskip1.5cm
{\large Yugo Abe\footnote{E-mail: 13st301k@shinshu-u.ac.jp }$^*$, 
Takeo Inami\footnote{E-mail: takeo.inami@riken.jp}$^{\dagger}$, 
Yoshiharu Kawamura\footnote{E-mail: haru@azusa.shinshu-u.ac.jp}$^*$ 
and Yoji Koyama\footnote{E-mail: ykoyama@phys.nthu.edu.tw}$^{\ddagger}$}

\vskip 1.5em
{\large\itshape $^*$Department of Physics, Shinshu University, Matsumoto 390-8621, Japan\\
\large\itshape $^{\dagger}$Mathematical Physics Lab., 
Riken Nishina Center, Saitama 351-0198, Japan\\
\large\itshape $^{\dagger}$Department of Physics, National Taiwan University, Taipei 10617, Taiwan, R.O.C.\\
\large\itshape $^{\ddagger}$Department of Physics, National Tsing Hua University, Hsinchu 30013, Taiwan, R.O.C.}  \vskip 4.5em
\end{center}

\begin{abstract}
We study the stabilization of an extra-dimensional radius in the presence of  a Wilson line phase
of an extra $U(1)$ gauge symmetry on a five-dimensional space-time,
using the effective potential relating both the radion and the Wilson line phase at the one-loop level.
We find that the radion can be stabilized by the introduction of a small number of fermions.
\end{abstract}

\end{titlepage}

\newpage

\abovedisplayskip=1.0em
\belowdisplayskip=1.0em
\abovedisplayshortskip=0.5em
\belowdisplayshortskip=0.5em
 
\parskip=0.5em
\parskip=0.25em

\section{Introduction}

Superstring theory is a powerful candidate of particle physics
including quantum gravity.
It is defined on the 10-dimensional (10D) space-time,
and necessarily possesses extra dimensions from the viewpoint of our 4D world.
Many scalar fields and extra gauge symmetries generally appear after compactification.
Scalar fields such as moduli and dilaton are massless at the tree level,
and how those fields are stabilized with suitable masses is a big issue.
Interesting solutions have been proposed in type IIB string theory~\cite{KKLT,CCQ}.

The study of effective field theories on a higher-dimensional space-time 
can provide useful hints
to the stabilization of massless scalar fields.

In a higher-dimensional gravity theory,
a scalar field called $\lq\lq$radion'' appears in the extra-dimensional components of graviton,
and its vacuum expectation value is related to the size of extra space.
The stabilization of the radius is crucial for the solution to the hierarchy problem 
in the Randall-Sundrum model~\cite{R&S}.
It is also vital for the realization of inflation based on the radion~\cite{Mazumdar}.
The radius is fixed through the interactions with bulk scalars 
in the Randall-Sundrum background
based on an orbifold $S^1/Z_2$~\cite{G&W,M&O}.
The stabilization is realized by the quantum effects of graviton and fermions
 in a 5D model with an extra space $S^1$~\cite{FIK}.
        
In a higher-dimensional gauge theory,
the extra-dimensional components of gauge bosons are massless at the tree level 
due to the gauge invariance,
their zero modes become dynamical degrees of freedom
called the Wilson line phases and are stabilized by quantum corrections~\cite{Hosotani}.
There is a possibility that realistic gauge symmetries 
including the standard model ones survive
after the stabilization of the Wilson line phases.
As a solution of the gauge hierarchy problem, 
it has been pointed out that the Wilson line phase receives 
finite radiative corrections on its mass
and can play the role of Higgs boson~\cite{HIL}.
An inflation model has been proposed based on the idea 
that the Wilson line phase becomes the inflaton~\cite{ACCR}.

Hence, it is interesting to study whether the radion is stabilized 
in the presence of  Wilson line phases or not.
This question has been examined in a specific gauge-Higgs unification model~\cite{M&S,S}. Their studies are based on the warped space-time with the Randall-Sundrum metric and the branes on the fixed points of $S^1/Z_2$. In addition to radiative corrections from the bulk gauge bosons, their brane-localized kinetic terms are crucial to stabilize the radion.

In this paper, we investigate how a Wilson line phase and the Casimir energy from various bulk fields involve the stabilization of the radion in a different setup. Concretely, we study the stabilization of an extra-dimensional radius of $S^1$
in the presence of  a Wilson line phase
of an extra $U(1)$ gauge symmetry on a 5D space-time with a flat background metric and without branes,
using the effective potential relating both the radion and the Wilson line phase 
at the one-loop level.

The content of our paper is as follows.
In the next section, we derive the effective potential 
including the radion and the Wilson line phase at the one-loop level.
In Sect. 3, we study the asymptotic behaviors of the effective potential,
and obtain the conditions to stabilize the radion.
It will be shown that incorporating neutral fermions is crucial to achieve the stability of the potential. 
In the last section, we give conclusions and discussions.


\section{One-loop effective potential}

We derive the effective potential concerning 
both the radion and the Wilson line phase at the one-loop level,
in the framework of  gravity theory coupled to a $U(1)$ gauge theory 
defined on a 5D space-time
including $S^1$ as an extra space. 

\subsection{Warm-up}

As a warm-up, we obtain the effective potential 
evaluating contributions from a massive complex scalar field $\varphi$
coupled to an extra $U(1)$ gauge boson $B_{M}$ in a 5D gravity theory.
The action of $\varphi$ is given by
\begin{eqnarray}
S_{\varphi}=-\int \sqrt{-\hat{g}_{5}} \left[\hat{g}^{MN} (D_M \varphi)^* (D_N \varphi) 
+ m_{\varphi}^2 |\varphi|^2\right] d^5x,
\label{S-varphi}
\end{eqnarray}
where $\hat{g}_{5}={\rm det}\hat{g}_{MN}$, 
$\hat{g}_{MN}$ ($M,N=0,1,2,3,5$) is the metric of 5D space-time,
$D_{M}=\partial_{M}-ig'_{5} q'_{\varphi} B_{M}$
and $m_{\varphi}$ is a mass of $\varphi$.
 $g'_{5}$ is the 5D gauge coupling constant 
and $q'_{\varphi}$ is the $U(1)$ charge  of $\varphi$.

We set $\hat{g}_{MN}$ as
\begin{eqnarray}
\hat{g}_{MN}=\Phi^{-1/3}\left(\begin{array}{cc}g_{\mu \nu}+A_{\mu}A_{\nu}\Phi 
& A_{\mu}\Phi\\ A_{\nu}\Phi & \Phi \end{array}\right) ,
\label{gMN}
\end{eqnarray}
then $\sqrt{-\hat{g}_{5}}=\Phi^{-1/3}\sqrt{-g_{4}}$.

We compactify the extra space on $S^{1}$ with the circumference $L=2\pi R$
and impose the periodic boundary conditions on every field.
Then, fields are given by the Fourier expansions
\begin{eqnarray}
&~& \hat{g}_{MN}\left(x^{\mu}, y\right)=\sum^{\infty}_{n=-\infty}\hat{g}^{(n)}_{MN}\left(x^{\mu}\right)e^{i\frac{2\pi n}{L}y},
\label{KK1}\\
&~& B_{M}\left(x^{\mu}, y\right)=\frac{1}{\sqrt{L}}\sum^{\infty}_{n=-\infty}B_{M}^{(n)}\left(x^{\mu}\right)e^{i\frac{2\pi n}{L}y},
\label{KK2}\\
&~& \varphi\left(x^{\mu}, y\right)=\frac{1}{\sqrt{L}}\sum^{\infty}_{n=-\infty}\varphi^{(n)}\left(x^{\mu}\right)e^{i\frac{2\pi n}{L}y},
\label{KK3}
\end{eqnarray}
where $x^{\mu}$ $\left(\mu=0,...,3 \right)$ and $y$ 
denote 4D coordinates and the $S^1$ one, respectively. 
We have two massless 4D scalar fields $\Phi^{(0)}$ and $B_{5}^{(0)}$ besides $\varphi^{(0)}$. 

We take the Minkowski metric $\eta_{\mu\nu}$ as the background value of the 4D metric 
$g_{\mu\nu}$, i.e., $\langle g_{\mu\nu}^{(0)}\rangle=\eta_{\mu\nu}$.
Other zero modes are assumed to have the following classical values, 
\begin{eqnarray}
\langle A_{\mu}^{(0)}\rangle =0,~~ 
\langle \Phi^{(0)}\rangle=\phi,~~ \langle B^{(0)}_{\mu}\rangle=0,~~ 
\langle B_5^{(0)}\rangle=\frac{\theta}{g'_{4}L},
\label{classical values}
\end{eqnarray}
where $\phi$ is the radion and $\theta$ is the Wilson line phase 
given by $\theta=g'_{4}\int^{L}_{0}dy\langle B_5^{(0)}\rangle$ and $g'_{4}=g'_{5}/\sqrt{L}$.
The massive modes called the Kaluza-Klein (KK) modes are assumed 
to have zero classical values.

To compute the one-loop potential, it suffices to know the quadratic
terms of $\varphi^{(n)}$ in the 4D action.
It is given by
\begin{eqnarray}
S_{\varphi}= \sum_{n=-\infty}^{\infty} \int \left[\varphi^{(n)*} 
\left\{- \raisebox{-1mm}{\LBox} + \left(\frac{2\pi n+q'_{\varphi}\theta}{\phi^{1/2} L}\right)^2 
+ \frac{m_{\varphi}^2}{\phi^{1/3}}\right\} \varphi^{(n)}\right] d^4x.
\label{S-varphi-4D}
\end{eqnarray}
Taking the standard procedure, the effective potential at the one loop level is given by
\begin{eqnarray}
V_{\varphi}^{\rm{eff}}=2 V^{\rm{eff}}(m_{\varphi}, q'_{\varphi}),~~
V^{\rm{eff}}(m, q') \equiv \frac{1}{2} \sum^{\infty}_{n=-\infty}\int 
\frac{d^{4}p_{\rm{E}}}{\left(2\pi \right)^{4}}
\ln\left[p^{2}_{\rm{E}}+\left(\frac{2\pi n+q'\theta}{\phi^{1/2} L}\right)^2
+\frac{m^2}{\phi^{1/3}} \right].
\label{Vvarphi}
\end{eqnarray} 
After the integration of the Euclidean momentum $p_{\rm{E}}$, 
$V^{\rm{eff}}(m, q')$ is calculated as
\begin{eqnarray}
&~& V^{\rm{eff}}(m, q') 
= - \frac{3}{4\pi^{2}}\frac{1}{\phi^{2}L^{4}}{\rm Re}
\biggl[{\rm Li}_{5}(e^{-Lm\phi^{1/3}}e^{iq'\theta})
+Lm\phi^{1/3}{\rm Li}_{4}(e^{-Lm\phi^{1/3}}e^{iq'\theta})
\nonumber \\
&~& ~~~~~~~~~~~~~~~~~~~~~~~~~~~~~~~~~~~~~~~~~~ 
\left. +\frac{1}{3}L^{2}m^{2}\phi^{2/3}{\rm Li}_{3}(e^{-Lm\phi^{1/3}}e^{iq'\theta}) \right] ,
\label{Veff}
\end{eqnarray}
where ${\rm Li}_{n}(x)$ are the polylogarithm functions
\begin{equation}
{\rm Li}_{n}(x)= \sum^{\infty}_{k=1}\frac{x^{k}}{k^{n}}  . \label{polylogarithm function}
\end{equation}

\subsection{Our model}

Our model consists of the 5D graviton $\hat{g}_{MN}$, 
a $U(1)$ gauge boson $B_{M}$, $c_{1}$ charged fermions $\psi_{i}$ ($i=1, \cdots, c_{1}$),
and $c_{2}$ $U(1)$ neutral fermions $\eta_{l}$ ($l=1, \cdots ,c_{2}$).
We take $M^4 \times S^1$ as a background 5D space-time
and impose the periodic boundary conditions on every field.
Here $M^4$ is the Minkowski space.

On ground of the Lorentz covariance and the gauge invariance
and consideration of the statistics for particles, 
the effective potential due to the one-loop of $X$ field is written as,
\begin{eqnarray}
V_{X}^{\rm{eff}}=f_{X} V^{\rm{eff}}(m_X, q'_X),
\label{VX}
\end{eqnarray}
where $f_X$, $m_X$ and $q'_X$ are the physical degrees of freedom, 
the mass and the $U(1)$ charge of $X$, respectively.
The values of $f_X$, $m_X$ and $q'_X$ for each field are given in Table \ref{T1}.
Note that the graviton and the $U(1)$ gauge boson own no $U(1)$ charge and are massless.
Unphysical modes of  $\hat{g}_{MN}$ and $B_M$ are eliminated 
by the contributions from their ghosts.
\begin{table}[hctb]
\caption{The physical degrees of freedom, mass and $U(1)$ charge for each field.}
\label{T1}
\begin{center}
\begin{tabular}{c|c|c|c} \hline
$X$ & $f_X$ & $m_X$ & $q'_X$ \\ \hline\hline
 $\hat{g}_{MN}$ & $5$ & $0$ & $0$ \\ \hline
$B_M$ & $3$ & $0$ & $0$ \\ \hline
$\psi_i$ & $-4$ & $m_i$ & $q'_i$ \\ \hline
$\eta_l$ & $-4$ & $\mu_l$ & $0$ \\ \hline
$\varphi$ & $2$ & $m_{\varphi}$ & $q'_{\varphi}$ \\ \hline
\end{tabular}
\end{center}
\end{table}

Hence, we obtain the full effective potential at the one-loop level~\cite{A&C,DPQ,P&P}
\begin{eqnarray}
V(\phi,\theta) &=& \sum_X V_{X}^{\rm{eff}}
 = 8 V^{\rm{eff}}(0, 0) -4\sum_i V^{\rm{eff}}(m_i, q'_i) 
 -4\sum_l V^{\rm{eff}}(\mu_l, 0) 
\nonumber \\
&=& -\frac{6}{\pi^{2}}\frac{1}{\phi^{2}L^{4}}\zeta(5)
\nonumber\\
&~&  +\sum_i \frac{3}{\pi^{2}}\frac{1}{\phi^{2}L^{4}}{\rm Re}
\biggl[{\rm Li}_{5}(e^{-Lm_i\phi^{1/3}}e^{iq'_i\theta})
+Lm_i\phi^{1/3}{\rm Li}_{4}(e^{-Lm_i\phi^{1/3}}e^{iq'_i\theta})
\nonumber \\
&~& ~~~~~~~~~~~~~~~~~~~~~~~ \left. +\frac{1}{3}L^{2}m_i^{2}\phi^{2/3}
{\rm Li}_{3}(e^{-Lm_i\phi^{1/3}}e^{iq'_i\theta})\right] 
\nonumber\\
&~&  +\sum_l \frac{3}{\pi^{2}}\frac{1}{\phi^{2}L^{4}}\biggl[{\rm Li}_{5}(e^{-L\mu_l\phi^{1/3}})
+L\mu_l\phi^{1/3}{\rm Li}_{4}(e^{-L\mu_l\phi^{1/3}})
\nonumber \\
&~& ~~~~~~~~~~~~~~~~~~~~~~~ \left. +\frac{1}{3}L^{2}\mu_l^{2}\phi^{2/3}
{\rm Li}_{3}(e^{-L\mu_l\phi^{1/3}}) \right],
\label{Vphitheta}
\end{eqnarray}
where we use the formula relating the Riemann zeta function $\zeta(n)$
\begin{equation}
\zeta(n)= \sum^{\infty}_{k=1}\frac{1}{k^{n}} = {\rm Li}_n(1). 
\label{zeta function}
\end{equation}

\begin{figure}[hctb]
\begin{center}
\includegraphics[width=120mm]{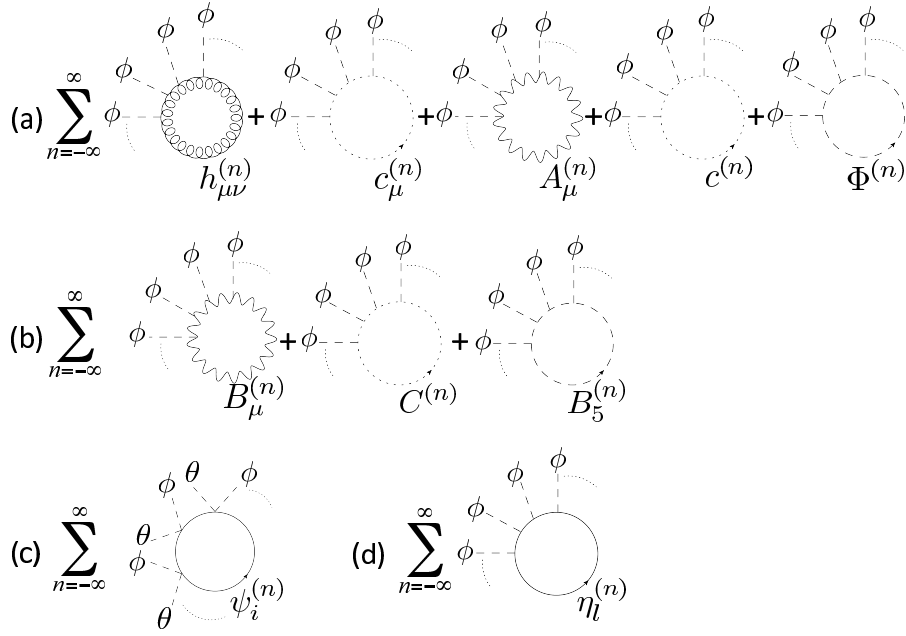}
\vspace{0mm}
\vskip-\lastskip
\caption{One-loop diagrams due to 5D graviton and its ghosts ($c_{\mu}$ and $c$), 
5D gauge boson and its ghost ($C$), charged fermions and neutral ones.}
\label{fig:one}
\end{center}
\end{figure}

The one-loop diagrams due to 5D graviton and its ghosts ($c_{\mu}$ and $c$), 
5D gauge boson and its ghost ($C$), charged fermions and neutral ones
are shown in Fig.\ref{fig:one} (a), (b), (c) and (d), respectively.
\vspace{0mm}

\section{Stabilization of radion}

We study the asymptotic behaviors of the effective potential and its stability,
in the simple case that
$\psi_i$ and $\eta_l$ have common masses $m$ and $\mu$, respectively,
and $\psi_i$ have a common $U(1)$ charge $q'_i = 1$.
Then, the potential is written by
\begin{eqnarray}
&~& V(\phi,\theta) 
=-\frac{6}{\pi^{2}}\frac{1}{\phi^{2}L^{4}}\zeta(5)\nonumber\\
&~&~~ +c_{1}\frac{3}{\pi^{2}}\frac{1}{\phi^{2}L^{4}}{\rm Re}\left[{\rm Li}_{5}(e^{-Lm\phi^{1/3}}e^{i\theta})
+Lm\phi^{1/3}{\rm Li}_{4}(e^{-Lm\phi^{1/3}}e^{i\theta})
+\frac{1}{3}L^{2}m^{2}\phi^{2/3}{\rm Li}_{3}(e^{-Lm\phi^{1/3}}e^{i\theta})\right] \nonumber\\
&~&~~ +c_{2}\frac{3}{\pi^{2}}\frac{1}{\phi^{2}L^{4}}\left[{\rm Li}_{5}(e^{-L\mu\phi^{1/3}})
+L\mu\phi^{1/3}{\rm Li}_{4}(e^{-L\mu\phi^{1/3}})
+\frac{1}{3}L^{2}\mu^{2}\phi^{2/3}{\rm Li}_{3}(e^{-L\mu\phi^{1/3}}) \right]. 
\label{Veff-full}
\end{eqnarray}

\subsection{Asymptotic behaviors}

First, we investigate the behavior of potential for large values of $\phi$.
For the region of $\phi$ satisfying 
\begin{eqnarray}
e^{-Lm\phi^{1/3}}, ~~ e^{-L\mu\phi^{1/3}}, ~~
{\rm Li}_{n}(e^{-Lm\phi^{1/3}}), ~~ {\rm Li}_{n}(e^{-L\mu\phi^{1/3}}) \ll 1,
\label{large-phi}
\end{eqnarray}
the potential can be approximated as
\begin{eqnarray}
V(\phi, \theta)\simeq-\frac{6}{\pi^{2}}\frac{1}{\phi^{2}L^{4}}\zeta(5). 
\label{phi's infinity contribution}
\end{eqnarray}
It approaches to zero from below as $\phi$ goes to infinity 
independently of  $\theta$ and other parameters. 
It is a natural consequence from the fact that the physical circumference of $S^{1}$ 
is given by $L_{\rm{phys}}=\phi^{1/3}L$
and the effective potential should vanish for $L_{\rm{phys}}\rightarrow\infty$ 
due to the 5D gauge invariance.

Next, we study the behavior of the potential for small values of $\phi$.
The polylogarithm functions can then be approximated as
\begin{eqnarray}
{\rm Li}_{n}\left[(1-Lm\phi^{1/3})e^{i\theta}\right]&\simeq& 
\sum^{\infty}_{k=1}\frac{(1-kLm\phi^{1/3})}{k^{n}}e^{ik\theta}
\nonumber\\
&=& {\rm Li}_{n}(e^{i\theta})-Lm\phi^{1/3}{\rm Li}_{n-1}(e^{i\theta}) 
\label{polylogarithm of small phi}
\end{eqnarray}
and
\begin{eqnarray}
{\rm Li}_{n}(1-L\mu\phi^{1/3})&\simeq&\zeta(n)-L\mu\phi^{1/3}\zeta(n-1).
\end{eqnarray}
Then, the potential takes the form
\begin{eqnarray}
V(\phi, \theta)&\simeq&
\frac{3}{\pi^{2}}\frac{1}{\phi^{2}L^{4}}\left[\left(c_{2}-2\right)\zeta(5)
-\frac{2c_{2}}{3}L^{2}\mu^{2}\phi^{2/3} \zeta(3)
-\frac{c_{2}}{3}L^{3}\mu^{3}\phi\zeta(2) \right. \nonumber\\
&&~~~~~~~~~~~ \left.+{\rm Re}\left\{c_{1}{\rm Li}_{5}(e^{i\theta})
-\frac{2c_{1}}{3}L^{2}m^{2}\phi^{2/3}{\rm Li}_{3}(e^{i\theta})
-\frac{c_{1}}{3}L^{3}m^{3}\phi{\rm Li}_{2}(e^{i\theta}) \right\}\right].
\label{effective potential when phi is so small 1}
\end{eqnarray}

For sufficiently small values of $\phi$, the mass independent terms are dominant
and almost determine the behavior of the potential near $\phi=0$.
Here, we consider two cases with $\theta = 0$ and $\theta = \pi$.

~~\\
(i) Case with $\theta=0$

\begin{figure}[hctb]
\begin{center}
\includegraphics[width=150mm]{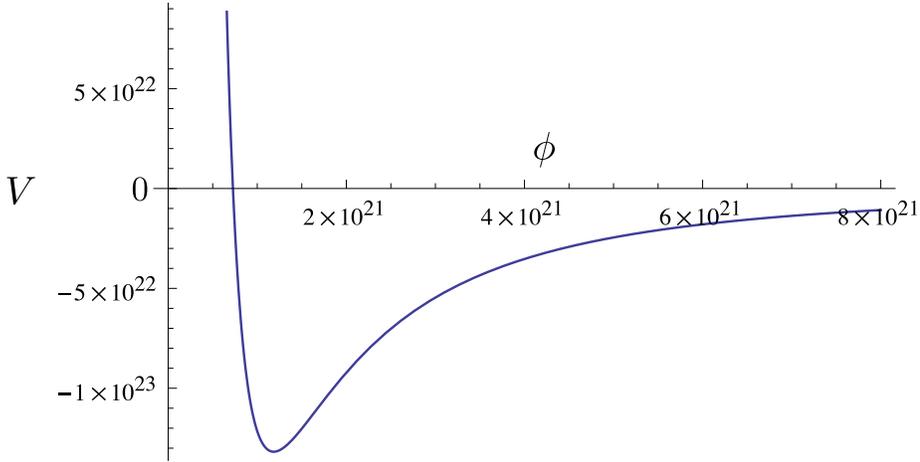}
\vspace{-40mm}
\vskip-\lastskip
\caption{The one-loop effective potential $V(\phi,\theta)$ for $\theta=0$ 
with $c_{1}=3$, $c_{2}=3$, $m=1\times 10^{10}${\rm GeV}, 
$\mu=1\times 10^{10}${\rm GeV} and $L=3\times 10^{-17}$${\rm GeV}^{-1}$. 
Here the unit of the longitudinal axis is ${\rm GeV}^4$. 
The same convention is used in Fig.\ref{fig:three}, Fig.\ref{fig:four} and Fig.\ref{fig:five}.}
\label{fig:two}
\end{center}
\end{figure}

In this case, the polylogarithm functions are just the $\zeta$-functions, i.e., 
${\rm Li}_{n}(e^{0})={\rm Li}_{n}(1)=\zeta(n)$, 
and positive constants for $n>1$. 
There are no differences between the charged fermions and neutral ones
in point of contributing to the potential. 
The potential is approximated as
\begin{eqnarray}
V(\phi, \theta=0)&\simeq&\frac{3}{\pi^{2}}\frac{1}{\phi^{2}L^{4}}
\left[\left(c_{1}+c_{2}-2\right)\zeta(5)
-\frac{2}{3}L^{2}\phi^{2/3}(c_{1}m^{2}+c_{2}\mu^{2})\zeta(3)\right.
\nonumber \\
&~& ~~~~~~~~~~~~ \left. -\frac{1}{3}L^{3}\phi(c_{1}m^{3}+c_{2}\mu^{3})\zeta(2)\right]
\end{eqnarray}
We see that the first term is dominant for $\phi\ll1$, 
and the potential is repulsive near the origin if $c_1 + c_2 > 2$, that is,
\begin{eqnarray}
c_1+c_2 \geq 3, \label{matter number condition}
\end{eqnarray}
where we use that $c_1$ and $c_2$ take zero or positive integers.
As $\phi$ becomes larger, i.e., $L^{3}m^{3}\phi\simeq 1$, 
the second and the third terms begin to exceed the first term, 
then the potential becomes negative and attractive. 
Taking into account the behavior of the potential for large value of $\phi$, 
we conclude that $V(\phi, \theta=0)$ has a minimum. 
A typical shape of $V(\phi, \theta=0)$ is shown in Fig.\ref{fig:two}. 
This result agrees with that in the previous works~\cite{A&C,P&P,R&R}.

~~\\
(ii) Case with $\theta=\pi$

In this case, the polylogarithm functions can be evaluated as follows.
\begin{eqnarray}
{\rm Li}_{n}(e^{i\pi})={\rm Li}_{n}(-1)=-1+\frac{1}{2^{n}}-\frac{1}{3^{n}}
+\frac{1}{4^{n}}- \cdots=\left(2^{1-n}-1 \right)\zeta(n).
\label{Lin(-1)}
\end{eqnarray}
Some sample values of ${\rm Li}_n(-1)$ are given in Table \ref{Table2}. 
${\rm Li}_{n}(-1)$ take negative values, 
and then flip the sign of the terms proportional to $c_{1}$ 
in (\ref{effective potential when phi is so small 1}).
\begin{table} [hctb]
\caption{Numerical values of ${\rm Li}_n(-1)$ for some values of $n$.}
\label{Table2}
\begin{center}
\begin{tabular}{c|ccccc} \hline
$n$ & 2 & 3 & 4 & 5 & 6 \\ \hline
${\rm Li}_n(-1)$ & $-0.822$ & $-0.90$ & $-0.947$ & $-0.96$ & $-0.986$ \\
\end{tabular}
\end{center}
\end{table}\\

The potential is given by
\begin{eqnarray}
V(\phi, \theta=\pi)&\simeq&\frac{3}{\pi^{2}}\frac{1}{\phi^{2}L^{4}}
\left[\left\{c_{2}-2-c_{1}\left(1-2^{-4} \right)\right\}\zeta(5)
-\frac{2c_{2}}{3}L^{2}\mu^{2}\phi^{2/3} \zeta(3)-\frac{c_{2}}{3}L^{3}\mu^{3}\phi\zeta(2) \right.
\nonumber\\
&&~~~~~~~~~~ \left.+\frac{2c_{1}}{3}L^{2}m^{2}\phi^{2/3}\left(1-2^{-2} \right)\zeta(3)
+\frac{c_{1}}{3}L^{3}m^{3}\phi\left(1-2^{-1} \right)\zeta(2)\right] . 
\label{theta = pi potential 2}
\end{eqnarray}

\begin{figure}[hctb]
\begin{center}
\includegraphics[width=130mm]{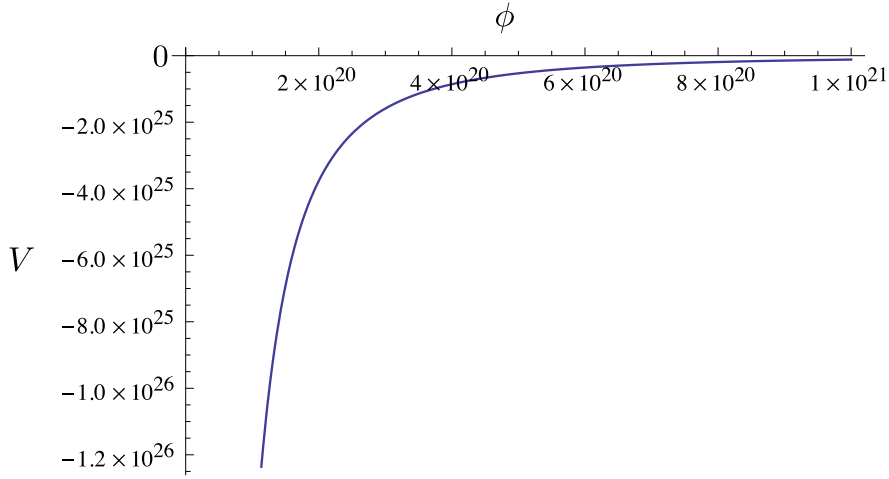}
\vspace{-20mm}
\vskip-\lastskip
\caption{$V(\phi, \theta)$ for $\theta=\pi$ with $c_{1}=3$, $c_{2}=0$, 
$m=1\times 10^{10}$GeV and $L=3\times 10^{-17}\rm{GeV}^{-1}$.}
\label{fig:three}
\end{center}
\end{figure}

In the absence of neutral fermions, i.e.,  $c_{2}=0$,
$V(\phi, \theta=\pi)$ is attractive near the origin and approaches 
to negative infinity for any $c_{1}$.
A typical shape of $V(\phi, \theta=\pi)$ is shown in Fig.\ref{fig:three}.

In the introduction of neutral fermions, the potential can have a finite minimum,
if the following condition is fullfilled
\begin{eqnarray}
c_{2}>2+c_{1}. \label{property of number of matter}
\end{eqnarray}
It comes from that the coefficient of the first term 
in $V(\phi, \theta=\pi)$ should be positive.
Note that the condition (\ref{matter number condition}) automatically holds 
if (\ref{property of number of matter}) is satisfied,
because $c_1 + c_2 > 2 + 2 c_1 \geq 2$.

\begin{figure}[hctb]
\begin{center}
\includegraphics[width=140mm]{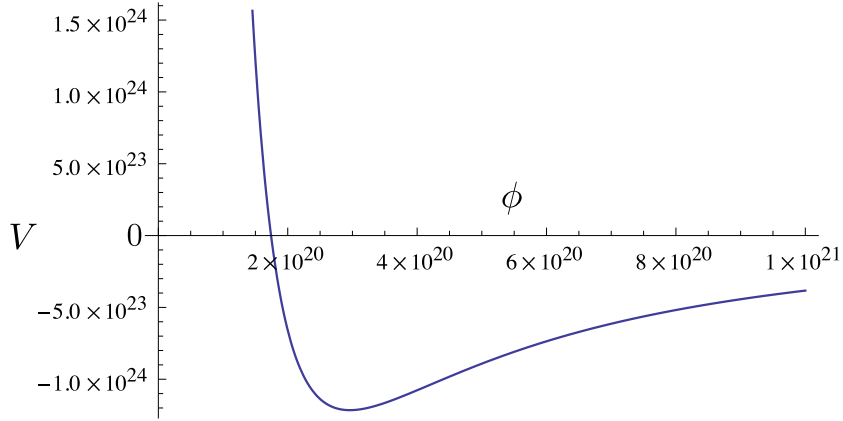}
\vspace{-40mm}
\vskip-\lastskip
\caption{$V(\phi, \theta)$ for $\theta=\pi$ with $c_{1}=1$, $c_{2}=4$, 
$m=1\times 10^{10}$GeV, 
$\mu=1\times 10^{10}$GeV and $L=3\times 10^{-17}\rm{GeV}^{-1}$.}
\label{fig:four}
\end{center}
\end{figure}

Fig.\ref{fig:four} shows a typical shape of the potential with a finite minimum
(\ref{property of number of matter}) being satisfied. 

\subsection{Stability}

First, we study the $\theta$ dependence of $V(\phi, \theta)$.
The $\theta$ dependent term in $V(\phi, \theta)$ has the form
\begin{eqnarray}
{\rm Re}{\rm Li}_n (e^{-Lm \phi^{1/3}} e^{i\theta}) = \sum_{k=1}^{\infty} 
\frac{\cos k\theta}{k^n} e^{-kLm \phi^{1/3}}.
\label{ReLin}
\end{eqnarray}
Its maximum and minimum with respect to $\theta$
for fixed values of $\phi$ can be found by computing its derivatives in $\theta$.
By noting
\begin{eqnarray}
\frac{d}{d \theta}{\rm Re}{\rm Li}_n (e^{-Lm \phi^{1/3}} e^{i\theta}) 
= -\sum_{k=1}^{\infty} \frac{\sin k\theta}{k^{n-1}} e^{-kLm \phi^{1/3}} =0
\label{dReLin}
\end{eqnarray}
and the second derivative of (\ref{ReLin})
\begin{eqnarray}
\frac{d^2}{d \theta^2}{\rm Re}{\rm Li}_n (e^{-Lm \phi^{1/3}} e^{i\theta}) 
= -\sum_{k=1}^{\infty} \frac{\cos k\theta}{k^{n-2}} e^{-kLm \phi^{1/3}},
\label{d2ReLin}
\end{eqnarray}
we find that the maximum and the minimum occur at $\theta = 0$ 
and $\theta = \pi$, respectively.

Hence, we have only to examine the behavior of $V(\phi, \theta)$ around $\theta = \pi$ 
for the stabilization of radius.
As seen in the previous subsection,
the potential has a finite minimum in the presence of neutral fermions 
whose numbers $c_2$ are bigger than $c_1 + 2$
and then the radion is stabilized at a certain finite value of $\phi$.

\begin{figure}[hctb]
\begin{center}
\includegraphics[width=140mm]{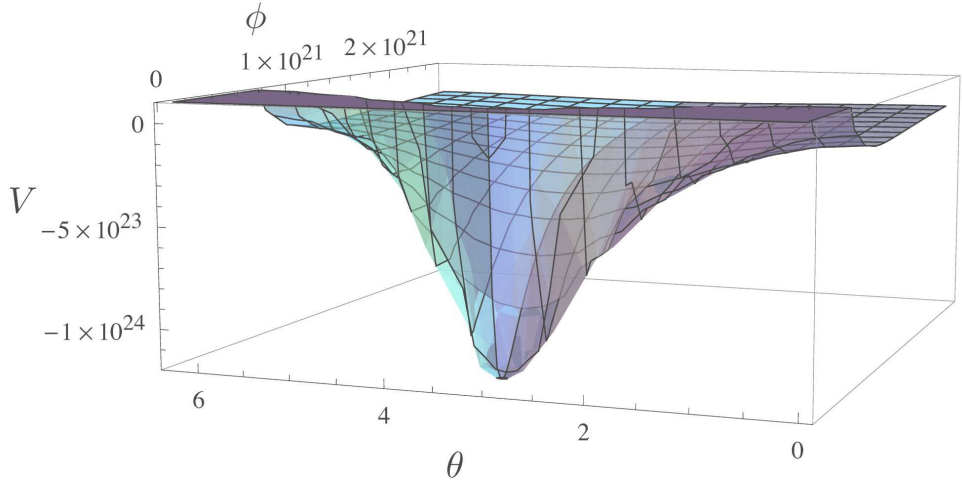}
\vskip-\lastskip
\caption{The potential $V(\phi, \theta)$ with $c_{1}=1$, $c_{2}=4$, 
$m=1\times 10^{10}$GeV, 
$\mu=1\times 10^{10}$GeV and $L=3\times 10^{-17}\rm{GeV}^{-1}$.}
\label{fig:five}
\end{center}
\end{figure}

A typical shape of $V(\phi, \theta)$ is depicted in Fig. \ref{fig:five}. 
From this figure, we see that the true minimum of the potential is located 
on the line of $\theta=\pi$.
The values of $\phi$ and the potential at the minimum depend on the parameters
such as $m$, $\mu$ and $L$. 
Their values do not modify the shape of the potential drastically,  
except for the cases such that $m\rightarrow\infty$, $\mu\rightarrow\infty$, 
$L=0$ and $L=\infty$.

Up to now, we have considered the case with massive fermions.
Here, we give a comment on the contributions from massless fermions. 
Their contributions have the same forms as those of the graviton and gauge boson, i.e., $1/\phi^2L^4$ 
up to their coefficients, 
and change the long distance (large $\phi$) behavior of the potential.
There are three cases such that i) $m=\mu=0$, ii) $m\neq 0$, $\mu=0$ 
and iii) $m=0$, $\mu\neq 0$. 
For the cases i) and ii), $V(\phi, \theta)$ does not have a stable minimum 
for any choice of the other parameters.
In contrast, the condition (\ref{property of number of matter}) is satisfied and
the potential has a stable minimum for the case iii).

\section{Conclusion and discussion}

We have studied the stabilization of  an extra-dimensional radius of $S^1$
in the presence of a Wilson line phase
of an extra $U(1)$ gauge symmetry on a five-dimensional space-time with a flat background metric and without branes,
using the effective potential relating the radion and the Wilson line phase at the one-loop level.
The Wilson line phases are incorporated in the introduction of charged fermions.
We have investigated the behavior of the potential for both large and small values of the radion,
and found that the potential does not have a finite minimum
in the case with only charged fermions as matter fields.
The stabilization of radion is realized in the presence of neutral fermions 
whose number is bigger than the number of charged ones by two.

Our result is different from those in a specific gauge-Higgs unification model~\cite{ M&S,S},
and this is mainly caused by the difference of setup. 
In the model with the Randall-Sundrum metric, the Casimir energy from the graviton is exponentially suppressed because the graviton is localized around the UV brane. The bulk gauge bosons spread over the bulk $S^1/Z_2$ and give sizable contributions to the potential, and then their gauge brane-localized kinetic terms are crucial to stabilize the radion. In contrast, we have considered the model with a flat background metric that every bulk field stretches out on $S^1$ and can give sizable contributions to the potential, and then the existence of bulk fermions is essential to the stabilization of the radion and the Wilson line phase.

It is straightforward to extend our analysis to 6D models.
The radion and the Wilson line phases in 6D models
are expected to have same properties as those in 5D ones, 
by imposing periodic boundary conditions on fields.

Subject left behind is to apply our potential $V(\phi, \theta)$ to an inflation model. 
By identifing the extra-dimensional component of the 5D gauge field 
and/or the scalar component of 5D metric as the inflaton, 
we examine whether the potential reproduces realistic inflation parameters or not. 
In other words, we have to study whether there is an allowed region of parameters 
in our potential to satisfy all the constraints, 
in order to realize the slow-roll inflation scenario~\cite{L&R}.
Now the radion and the Wilson line phase may compete in giving rise to inflation. 
Because the property of the radion differs from that of the Wilson line phase,
there is an interesting possibility that a hybrid inflation~\cite{Linde:hybrid} occurs
besides the extranatural inflation~\cite{ACCR} and the radion inflation~\cite{FIK}.
We are interested in which scalar dominates the energy density of the universe 
and then can play the role of inflaton. 
 
\section{Acknowledgement}
This work is supported in part by scientific grants from the Ministry of Education, Culture,
Sports, Science and Technology under Grant Nos.~21244036 and 20012487 (T.Inami),
Grant Nos.~21244036 and 22540272 (Y.Kawamura), 
and National Science Council under  Grant No.NSC-101-2112-M-007-021(Y.Koyama).
Y.A. benefitted much from his visit to National Taiwan University. Hewishes to thank Professor Ho to give him the chance of this visit.
T.I. wishes to thank CTS of NTU for a partial support.

\end{document}